**Investigating the Feasibility of Virtual Reality for Emotion Regulation with Youth**

**Background:** In the last ten years, Human-Computer Interaction (HCI) has seen a rise in research on how features of interactive technologies can be designed to enhance emotion regulation training [1]–a response in part due to the increase of mental health issues. Youth are a vulnerable population at risk for mental health issues with long wait times to get help [2]. Emotion regulation, the ability to modulate the intensity and duration of emotional states, has been explored as a trans-diagnostic intervention to improve mental health [3]. Youth aged 12-15 use less adaptive or maladaptive emotion regulation strategies (e.g., suppression) that can contribute to depression or anxiety disorders. Thus, training adaptive emotion regulation skills is crucial for this age group [4]. One of the most effective emotion regulation strategies for youth is cognitive reappraisal–changing how we think about a situation in order to decrease its emotional impact [5]. However, learning is difficult because it requires: (1) generation of meaningful emotionally laden social situations; (2) real-time feedback of internal psychological and physiological states; and (3) practice modifying the interpretation of multiple situations [6].

Virtual Reality (VR) is a computer-generated 3D environment that allows the user to experience a simulated world through stereoscopic 360 visuals, stereo audio, and 3D interaction with tracking sensors. The affordances of VR have been shown to strongly affect human emotional responses and interpretations of social situations when used in other intervention contexts [7]. As such, VR may provide unique opportunities for cognitive reappraisal skills development through three mechanisms: <u>Mechanism #1: virtual environment</u>–a simulated world that evokes the visceral experience of a realistic emotional response; <u>Mechanism #2: interaction & feedback</u>–possibilities to control and modify emotionally evocative aspects of the virtual environment; <u>Mechanism #3: taking multiple perspectives</u>–the ability to put the user in another's shoes.

As an example, you put on a VR headset and find yourself walking down the halls on the first day of high school; there are lots of new faces and you hope to make a good first impression <u>(Mech #1)</u>. You hear the sound of your heart beating and your breathing getting heavier and uneven, and other kids laughing–are they laughing at you <u>(Mech #2)</u>? Luckily, you have a super power that allows you to embody other peoples' perspectives <u>(Mech #3)</u>. You take the VR controller, point it at the older kid who was laughing, and teleport into their virtual body <u>(Mech #2)</u>. Suddenly, you can see yourself from their perspective and hear their thoughts–"look at all these new kids. I remember my first day I tripped down the stairs and broke a tooth. So embarrassing!" <u>(Mech #3).</u> You teleport back to your body and hear your heart rate drop to indicate that you have regulated the anxiety response <u>(Mech #2)</u>. They weren't laughing at you, but empathizing with your experience. Seeing and understanding this new perspective is cognitively reappraising the event. In this scenario, VR offers youth a way to practice cognitive reappraisal in a safe, meaningful, and emotionally laden environment with real-time feedback; thus meeting the reappraisal requirements (1-3) listed above, which are normally difficult to meet.

**Objectives & Research Questions:** There is a rise in VR's accessibility and popularity among youth (e.g., Oculus Quest 2 VR headset does not require a computer, costs $400 CAD, and has sold over 5 million units). My expertise in technology development of VR systems and content, design for mental health, and field evaluation skills make me well positioned to take this unique opportunity to investigate the overarching research question: **can the above three VR mechanisms (Mechanisms #1-3) address the three learning requirements of cognitive reappraisal skill development in youth aged 12-15?**

**Methodology:** I will conduct research that investigates the following **research questions** that derive from my objectives above: (1) <u>Requirements:</u> What are the minimum technical and design requirements for a VR platform to help youth learn and practice emotion regulation? (2) <u>Usability:</u> What are the system and interface usability factors that youth suggest for the VR prototypes? (3) <u>Feasibility:</u> What is the feasibility of using VR's three mechanisms to address the three learning requirements of cognitive reappraisal skill development during a six-week deployment with youth in their homes? (4) <u>Efficacy:</u> What are the subjective indicators of effects on cognitive reappraisal skills development (whether positive or negative) reported by youth? I will answer these research questions in three phases.

Phase 1 (Year 1) Develop a proof of concept prototyping platform: I will integrate my previous PhD work on technology-mediated emotion elicitation with cognitive reappraisal **learning requirements 1-3** (listed above) in order to ground the design of a platform for prototyping a series of VR experiences for cognitive reappraisal that are relevant for youth, e.g., first day at school simulation. I will develop 3D content and animations with Maya and Mixamo. I will custom-build design tools in Unity using C# and Unity's XR Plugin Management for an immersive standalone VR technology (e.g., Oculus Quest 2). These design tools will allow users to select and modify from a virtual menu of design features (e.g., aesthetic features and reference frame) and interaction mechanisms (e.g., locomotion or personalized and adaptive feedback) related to VR **mechanisms #1-3** and see their changes in real-time.

Phase 2 (Year 1 & 2) Conduct user-centered iterative design with youth: I will run an in-person two-week workshop with 12 youth aged 12-15 (balanced in regards to sex and gender) to experience and give iterative feedback on my initial prototypes and consider how they might customize VR **mechanisms #1-3** to their needs. I will recruit participants through my proposed supervisor Dr. Antle's connections to existing research sites. I will collect data on youth's ideas for customizations regarding VR **mechanisms #1-3** through video recordings (observations on their feature and interaction choices), and surveys (self-reported preferences). I will evaluate usability with youth-centric versions of traditional usability metrics including error analysis, learning time, System Usability Scale [8], and semi-structured interviews. Based on these results analyzed through statistical, descriptive and thematic analysis, I will iteratively revise and tune the VR prototypes to meet the needs of youth.

Phase 3 (Year 2) Deploy prototype with evaluation on feasability: I will use a waitlist controlled field study with 276 youth aged 12-15 (balanced in regards to sex and gender) using the youth-specific VR prototype deployed over six weeks. I will recruit youth through the Committee of Children (CfC) and who score ≥10 on the (emotion) Strengths and Difficulties Questionnaire [9] to increase meaningful impact. I will collect cognitive reappraisal data from pre- and post-assessment tests using a combination of the Emotion Regulation Questionnaire [10], semi-structured interviews, and logs collected automatically by the prototype that records the VR screen, user movement, and controller interactions. I will corroborate statistical analysis with interviews, self-report and observational data to assess whether VR's three mechanisms for addressing the three learning requirements of cognitive reappraisal skill development are feasible, and if so what are the important design features for other developers.

**Significance to NSE:** My work has six contributions: (1) technical development of a VR prototyping platform for other HCI researchers and/or psychologists to develop individualized cognitive reappraisal training scenarios, (2) digital content for a youth-specific simulated emotion regulation training environment, (3) design methodology on how to conduct embodied workshops with youth in VR, (4) design guidance, (5) empirical results on the feasibility of emotion regulation training through VR-mediated interventions for youth, and (6) an improved understanding, assessment, and training of cognitive reappraisal for youth. Providing youth with an effective way to regulate their emotions can improve their mental health, which will lead to improvements in education, socio-emotional, and economic outcomes for youth in Canada and globally.